\documentclass{JHEP3}
\usepackage{amssymb,amsfonts,amsmath,amsthm,latexsym,graphicx,verbatim,epsfig}
\newcommand{\be}{\begin{equation}}
\newcommand{\ee}{\end{equation}}
\newcommand{\bea}{\begin{eqnarray}}
\newcommand{\eea}{\end{eqnarray}}
\newcommand{\la}{\lambda}
\newcommand{\rcl}{\rho_{cl}}
\newcommand{\pd}{\partial}
\newcommand{\pdt}{\pd_{\tau}}

\title{\center{On evolutionary processes in large N quantum mechanics}}

\author{
    Vadim Asnin\footnotemark[1]\\
    Racah Institute of Physics\\
    Hebrew University \\
    Jerusalem 91904,
    Israel\\

    \footnotemark[1] {\tt vadim.asnin@mail.huji.ac.il\\}

        }

\abstract{Equations of motion of large N quantum mechanics are
solved for infinite N in the case of unbroken global O(N) symmetry.
It is shown that the only correction to the action is a change in
the potential. All characteristics of the motion (frequencies of
oscillations, times of rolling, a phase portrait etc.) can be
computed by means of a classical mechanics for the corrected
potential. In particular, a tunneling in the system is absent.}
\begin{document}

\section{Introduction}
Evolutionary processes have been studied in the framework of Quantum
Mechanics from its early days. Even the first complete formulation
of principles of Quantum Mechanics described a way to find a time
dependence of observables (Heisenberg eqs). In Quantum Field Theory
the common lore concerns mainly properties of vacua and
perturbations around them. Still, time dependent processes are
crucial in various branches of QFT (inflationary cosmology, QFT in
curved spaces etc). These processes include a rolling of a quantum
field from a maximum of a potential, oscillations around a minimum
and a tunneling from one minimum of a potential to others.

In general, a study of time-dependent processes in QFT is
prohibitively difficult. It can be carried out, however, in large
$N$ vector models.

A large $N$ vector model with a global $O(N)$ symmetry is a rare
example of a field theory where an exact vacuum state can be found
(for a review of large $N$ models see \cite{Moshe:2003xn}). This
model has served as a toy model for various problems common in QFT.
A three-dimensional scale-invariant model was shown to be quantumly
conformal \cite{Bardeen:1983rv} and to possess a non-trivial phase
structure \cite{Bardeen:1983st}.

General eqs. which govern time-dependent processes in the large $N$
vector model were suggested in \cite{Asnin:2009bs} and applied there
to the case of $d=3$. The applications included an exact rolling
solution of a $\phi^6$ model with vanishing energy and an
approximate solutions for oscillations around minima in a general
potential. In all cases the quantum processes were found to occur
faster than their classical counterparts, signalizing a presence of
terms with time derivatives in addition to a known correction to a
classical potential. A possible tunneling in the model was also
studied and the tunneling amplitude was found to be larger than in
the semiclassical computation.

The main difficulty in solving for time dependent processes in the
large $N$ model, as discussed in \cite{Asnin:2009bs}, is a necessity
to find a Green's function of a differential operator which itself
contains an unknown function. In this paper we show that this
problem can be solved in the case of $d=1$, which corresponds to a
large $N$ quantum mechanics.

A large $N$ quantum mechanics has been studied from various point of
view. Numerous approximations have been used in order to derive
systematically a $1/N$ expansion (for a review and a comparison of
some approximation technics see \cite{Mihaila:2000sr}, in
\cite{Ryzhov:2000fy} yet another approach is suggested). A
supersymmetric version of the large $N$ quantum mechanics was also
studied (for improvements that supersymmetry brings to the large $N$
expansion see \cite{Cooper:1994eh}). A certain supersymmetric
version of the large $N$ quantum mechanics was proposed as a
description of a D-brane probe in the bubbling supertubes
\cite{Imamura:2005nx}.

In this paper we solve EOM's derived in \cite{Asnin:2009bs} for the
quantum mechanics in the case of unbroken global $O(N)$ symmetry. We
show that the expectation value of the quantum field changes in time
according to a classical EOM with a modified potential, which means
that the effective action in this case does not involve any
corrections to the kinetic term, the only correction being therefore
that to the potential. All characteristics of the motion (like
frequencies of oscillations around minima of the effective potential
and rolling times in that potential, and possible types of the
motion in general) can be derived by means of the classical
mechanics. In addition, a tunneling is completely suppressed.

The paper is organized as follows. In section \ref{useful_formulas}
we review the large $N$ models in general and introduce eqs. that
govern a time evolution of the quantum system. In section
\ref{Classical_analysis} we rewrite a classical EOM in a form which
is convenient for a comparison with the quantum case. In section
\ref{Quantum_case} we solve the quantum EOM's. Section \ref{Summary}
is a summary of the results. We end up with an appendix
\ref{Tunneling} where we show that there is no tunneling in the
system in the limit $N\to\infty$ and confirm this result by
conventional methods of quantum mechanics.
\section{Scalar model in the large N limit - A review}\label{useful_formulas}

Let us consider an $O(N)$-symmetric Euclidean action for an $N$ -
component scalar field $\vec{\phi}$ in $d$ space-time dimensions

\be
S\left( \vec{\phi}\right) =\int \left[ \frac{1}{2}\left(
\partial _{\mu }\vec{\phi}\right) ^{2}+NU\left( \frac{
\vec{\phi}^{2}}{N}\right) \right] d^{d}x \, .\label{scalaraction}
\ee The potential has a Taylor expansion of the form \be U\left(
\frac{\vec{\phi}^{2}}{N}\right)
=\sum_{n=1}^{\infty}\frac{g_{2n}}{2n}\left(
\frac{\vec{\phi}^{2}}{N}\right) ^{n} \, ,
\label{ExampleOfPotential}\ee with $g_{2n}$ kept fixed as
$N\to\infty $. The corresponding partition function is \be Z=\int
D\vec{\phi}\,e^{-S(\vec{\phi})}\, . \ee In order to use the fact
that $\vec{\phi}$ has many components insert the following
representation of unity into $Z$:

\be 1\sim\int D\rho\, \delta (\vec{\phi}^{2}-N\rho)\sim \int D\rho
D\lambda\, e^{-i\int \frac{\lambda}{2}(\vec{\phi} ^{2}-N\rho
)d^{d}x}\, . \label{identity} \ee Now one can integrate over
$\vec{\phi}$ and obtain \be Z= \int D\rho D\lambda\,
 e^{-N S_{eff}(\rho,\,\la) }\, ,
\label{partition} \ee where
\begin {equation}
 S_{eff}(\rho,\la)=\frac{1}{2}\int\left[2U(\rho) -i\lambda\rho\right]d^{d}x
 +\frac{1}{2}Tr\ln \left( -\square +i\lambda \right)  \, .
 \label{Seff}
\ee

When $N$ is large, this form of the path integral suggests to use
the saddle point method to calculate the integral. The two saddle
point equations obtained by varying the auxiliary fields $\rho$ and
$\lambda $ are\footnote{We use here the definition $Tr=\int
d^Dx\,tr$}

\be 2U'(\rho)=i\lambda,\qquad \rho=tr\frac{1}{-\square+i\lambda}\, ,
\label{scalar_gap} \ee

This form of the saddle point equations is convenient in the case of
constant fields. These constant solutions can be found as extrema of
the effective potential \be
U_{eff}(\rho)=U(\rho)+\frac{2-d}{2\,d}\,\Gamma\left(1-\frac
d2\right)^{-\frac{2}{d-2}}\,\left(4\pi\rho\right)^{\frac{d}{d-2}}\,
,\label{Effective_Potential}\ee together with the following value of
the field $\la$: \be
i\la=\left[\frac{(4\pi)^{d/2}\rho}{\Gamma\left(1-\frac
d2\right)}\right]^{\frac{2}{d-2}}\,.\ee The necessity of the second
term in (\ref{Effective_Potential}) is most clearly seen in $d=1$,
where the theory becomes a quantum mechanics. In this case the
effective potential is \be U_{eff}(\rho)\Bigl|_{d=1}=
U(\rho)+\frac{1}{8\rho}\,.\label{1d_eff_potential}\ee Then, consider
the case of $U=\frac{g_2}{2}\rho$, which corresponds to a system of
$N$ decoupled harmonic oscillators. The last term in
(\ref{Effective_Potential}) shifts a minimum of the potential from
$\rho=0$ to a correct value $\rho=\frac{1}{2\sqrt{g_2}}$. Therefore
this term takes into account a spreading of the ground state wave
function.

If the fields are not constant then, as shown in
\cite{Asnin:2009bs}, the correct form of the equations is \be
2U'\bigl(\rho(x)\bigr)=i\lambda(x),\qquad \rho(x)=G(x,x),\qquad
\Bigl(-\Box_x+i\la(x)\Bigr)\,G(x,y)=\delta(x-y)\,\label{General_equations}\ee
where $\Box_x$ is a Laplacian w.r.t. $x$.

A major role in the above equations in played by the Green's
function $G(x,y)$. The field $\rho$ is equal to this function at
coincident points. In QFT this is divergent and is to be
regularized. We will concentrate on the case of $d=1$ where no
regularization is needed.

The main difficulty in solving equations (\ref{General_equations})
stems from the fact that $G(x,y)$ is a Green's function of the
operator which involves an unknown function $\la(x)$. In
\cite{Asnin:2009bs} there was found a particular exact solution of
these equations for $d=3$ and a potential
$U(\rho)=\frac{g_6}{6}\,\rho^3$, which corresponds to a model
$\vec{\phi}^6$. The particular solution found there corresponds to a
vanishing energy. In that case there is no scale in the problem
since the coupling constant $g_6$ is dimensionless in $d=3$.
Therefore the form of the solution is determined by dimensional
analysis up to dimensionless multiplicative constants which can be
fixed.\footnote{In \cite{Asnin:2009bs} there were also considered
approximate solutions of equations (\ref{General_equations})} In
this paper we solve the equations (\ref{General_equations}) in the
case of $d=1$.
\section{Classical analysis}\label{Classical_analysis}
In this section we consider a classical version of the large $N$
quantum mechanics. Our main purpose is to write the EOM in a form
that will be convenient for a comparison with its quantum
counterpart. For the same reason we work in the Euclidean signature.
We denote the Euclidean time by $\tau$.

The Euclidean Lagrangian of the system is \be
L=\frac12\left(\pd_{\tau}\vec{\phi}\right)^2+NU\left(\frac{\vec{\phi}^2}{N}\right)\,
.\ee We consider a phase with an unbroken global $O(N)$ symmetry,
which means that the field configurations we are interested in are
of the form \be \vec{\phi}(\tau)=\phi(\tau)\,\bigl(1,1,\ldots
1\bigr)\, ,\ee where the vector on the RHS has $N$ components. For
such field configurations the Lagrangian can be rewritten as \be
L=N\,\Bigl(\frac12\dot{\phi}^2+U(\phi^2)\Bigr)\,,\ee where a dot is
a derivative w.r.t. $\tau$. The Euclidean EOM is \be
\ddot{\phi}=\frac{dU}{d\phi}\,.\ee

In order to make this equation resemble the quantum one we introduce
a classical analog of the field $\rho$:\be
\rcl=\frac{\vec{\phi}^2}{N}=\phi^2\,.\ee Then, using the fact that
$\frac{dU}{d\phi}=2\phi\frac{dU}{d\rho}$ one can write a
differential equation of the third order for $\rcl$:\be
\dddot{\rho}_{cl}=8U'(\rcl)\,\dot{\rho}_{cl}+4U''(\rcl)\,\rcl\,\dot{\rho}_{cl}\,.\label{Classical_eq}\ee
This equation should be supplemented by initial conditions. Suppose
that at $\tau=0$ the field $\phi$ is equal to $\phi_0$ and its time
derivative is $\dot{\phi}_0$. Then the initial conditions for eq.
(\ref{Classical_eq}) are \be \rcl(0)=\phi_0^2,\qquad
\dot{\rcl}(0)=2\,\phi_0\,\dot{\phi}_0,\qquad
\ddot{\rcl}(0)=U'\left(\phi_0^2\right)+2\,\dot{\phi}_0^2\,.\ee This
is the EOM in the form of (\ref{Classical_eq}) that will be compared
to the quantum one in the next section.
\section{Quantum solution}\label{Quantum_case}
In this section we derive the quantum EOM for the field $\rho$
starting from general equations (\ref{General_equations}). The main
result is that the quantum EOM is again of the form
(\ref{Classical_eq}), but the initial conditions are different. It
will also be shown that a difference in the initial conditions can
be turned into a correction to a potential.

In the case of quantum mechanics eqs. (\ref{General_equations})
reduce to \be 2U'\bigl(\rho(\tau)\bigr)=i\lambda(\tau),\qquad
\rho(x)=G(\tau,\tau),\qquad
\Bigl(-\pdt^2+i\la(\tau)\Bigr)\,G(\tau,\tau_0)=\delta(\tau-\tau_0)\,\label{General_equations_1d}\,,\ee
and there is no need in a regularization.

We are going to consider a rolling of the system from a top of the
potential. Therefore we assume that the fields $\rho$ and $\la$
possess limits when $\tau\to\infty$. We will denote the limit of
$i\la$ by $m^2$ and assume in what follows that its value is
positive. Then we redefine the field $\la$:\be i\la=i\la_1+m^2\,
,\ee so that the new field $\la_1$ goes to 0 at infinity. In terms
of this new field the eqs. (\ref{General_equations_1d})
become \bea 2U'\bigl(\rho(\tau)\bigr)&=&i\lambda_1(\tau)+m^2\label{Rho_eq},\\
\rho(x)&=&G_m(\tau,\tau),\\
\Bigl(-\pdt^2+i\la(\tau)+m^2\Bigr)\,G_m(\tau,\tau_0)&=&\delta(\tau-\tau_0)\,\label{General_equations_m}\,,\eea
where the subscript of $G$ indicates the value of the parameter $m$
at which the Green's function is computed.

In one dimension the Green's function is closely related to
solutions of the corresponding homogeneous equation \be
\Bigl(-\pdt^2+i\la_1(\tau)+m^2\Bigr)\,g(\tau)=0\,.\label{Homogeneous_eq}\ee
Since $\la_1$ vanishes at infinity the solutions of this equation
look at infinity as $e^{\pm m\tau}$ (this approach is similar to
that used in \cite{Feinberg:1994qq}). We choose two linear
independent solutions $g_{1,2}$ of the homogeneous equation so that
the function $g_1(\tau)$ goes to 0 as $\tau\to-\infty$ and
$g_2(\tau)$ goes to 0 as $\tau\to+\infty$. We will fix their
normalization later.

In terms of these solutions the Green's function can be written as
\be G_m(\tau, \tau_0)=\Biggl\{\begin{array}{c}
                        \alpha_1\,g_1(\tau),\qquad \tau<\tau_0 \\
                        \alpha_2\,g_2(\tau),\qquad \tau>\tau_0
                      \end{array}\ee
where the coefficients $\alpha_{1,2}$ satisfy \be
\alpha_1\,g_1(\tau_0)=\alpha_2\,g_2(\tau_0),\qquad
\alpha_1\,g'_1(\tau_0)-\alpha_2\,g'_2(\tau_0)=1\,.\ee Solving these
equations leads to the following expression for the Green's
function: \be G_m(\tau,
\tau_0)=\frac{1}{W(g_1,g_2)}\,\Biggl\{\begin{array}{c}
                        g_1(\tau)\,g_2(\tau_0),\qquad \tau<\tau_0 \\
                        g_1(\tau_0)\,g_1(\tau),\qquad \tau>\tau_0
                      \end{array}\ee
where $W(g_1,g_2)\equiv g'_1(\tau)g_2(\tau)-g'_2(\tau)g_1(\tau)$ is
a Wronskian of the two solutions. It is independent of $\tau$ and we
fix the normalization of the basic solutions $g_{1,2}(\tau)$ so that
$W(g_1,g_2)=1$. With this normalization \be G_m(\tau,
\tau_0)=\Biggl\{\begin{array}{c}
                        g_1(\tau)\,g_2(\tau_0),\qquad \tau<\tau_0 \\
                        g_1(\tau_0)\,g_1(\tau),\qquad \tau>\tau_0
                      \end{array}\ee
and the Green's function with coincident points (which is equal to
$\rho$) is \be
G_m(\tau,\tau)\equiv\rho(\tau)=g_1(\tau)g_2(\tau)\,.\ee Using eq.
(\ref{Homogeneous_eq}) we write the equation for $\rho$:\be
\dddot{\rho}=4\bigl(i\la_1+m^2\bigr)\,\dot{\rho}+2\,i\dot{\la}\,\rho\,.\ee
Now, using (\ref{Rho_eq}) we get \be
\dddot{\rho}=8U'(\rho)\,\dot{\rho}+4U''(\rho)\,\rho\dot{\rho}\,.\label{Quantum_eq}\ee
This equation precisely coincides with the classical equation
(\ref{Classical_eq}). However, this does not mean that the possible
motions of the system are the same, since we need to specify initial
conditions. Eqs. (\ref{Classical_eq}, \ref{Quantum_eq}) are of the
third order and should be integrated once to be brought to a usual
form of equations of dynamics.

In order to integrate once the eq. (\ref{Quantum_eq}) define a new
field in analogy with the classical case:\be \Phi=\sqrt{\rho}\,.\ee
In terms of this new field a first integral of eq.
(\ref{Quantum_eq}) can be written as \be
\ddot{\Phi}=\frac{dU}{d\Phi}+\frac{\alpha}{\Phi^3}\,.\label{Quantum_Phi_eq}\ee
Here $\alpha$ is an integration constant. In order to fix it
consider constant solutions of (\ref{Quantum_Phi_eq}). They are
given by extrema of the function \be
\bar{U}(\Phi^2)=U(\Phi^2)-\frac{\alpha}{2\Phi^2}\,,\ee or if we
rewrite it in terms of $\rho$, by extrema of \be
\bar{U}(\rho)=U(\rho)-\frac{\alpha}{2\rho}\,.\ee We see that the
function $\bar{U}$ is of the form of a general effective potential
of the large $N$ vector model at $d=1$, which is given in
(\ref{1d_eff_potential}). This fixes $\alpha=-1/4$, and the
effective potential becomes \be
U_{eff}(\Phi^2)=U(\Phi^2)+\frac{1}{8\Phi^2}\,.\label{Quantum_effective_potential}\ee
The quantum EOM can be written in terms of the field $\Phi$ as \be
\ddot{\Phi}=\frac{dU(\Phi^2)}{d\Phi}-\frac{1}{4\Phi^3}\,,\ee or,
Wick rotating back to the Lorentzian time $t$, \be
\ddot{\Phi}=-\frac{dU(\Phi^2)}{d\Phi}+\frac{1}{4\Phi^3}\,,\label{Quantum_Phi_eq_fixed_const}\ee
This equation by construction correctly reproduces the extrema on
the quantum effective potential (\ref{1d_eff_potential}). Other
solutions of this equation describe time-dependent processes in the
system.

We have proven that the quantum mechanical EOM for an expectation
value of the field $\Phi$ (or $\rho$) is of a form of the classical
EOM but with a corrected potential. Therefore a classical intuition
about a motion of the system applies directly to the quantum case.
For example, a frequency of small oscillations around a minimum of
the effective potential is given by
$\omega^2=U_{eff}^{\prime\prime}(\Phi_0)$, where $\Phi_0$ is an
expectation value of the field $\sqrt{\rho}$ at the minimum of
$U_{eff}$. One can also draw a phase portrait on the phase plane
which would describe qualitatively all possible motions of the
system. In addition, the quantum mechanical tunneling is not
allowed. This is a consequence of the large $N$ limit. We show this
in a different way in appendix \ref{Tunneling}, where we also show
that the tunneling amplitude goes to zero as
$\pi^{N/2}\left[(N/2)!\right]^{-1}$.

\section{Summary and discussion}\label{Summary}
In this paper we considered the large $N$ quantum mechanics with an
unbroken global $O(N)$ symmetry. We found that the mean value of the
square of the field satisfies a classical EOM with modified
potential (\ref{Quantum_effective_potential}). Since the EOM is
purely classical we conclude that the correction to the potential is
the only difference between the classical action and the 1PI
effective action which governs the quantum evolution. In particular,
there are no terms with time derivatives, contrary to the case of
$d=3$, where such terms are present \cite{Asnin:2009bs}. This result
allows one to compute all characteristics of time-dependent
processes in an arbitrary potential (frequencies of oscillations
around minima of the potential, characteristic times of rolling, the
phase portrait etc) by the conventional means of the classical
mechanics. It also follows that the tunneling in the system is
suppressed in the large $N$ limit.

A desirable continuation of this work is to find a way to solve the
quantum EOM's in the case of higher dimensionality. This is much
more difficult, however. Although a reduction of the problem to a
one-dimensional case is always possible (assuming that the solution
we are looking for depends only on time), an expression for the
field $\rho$ in terms of solutions of a homogeneous eq. similar to
(\ref{Homogeneous_eq}) will in general involve an integral over a
mass, a fact that significantly complicates a computation.
\section*{Acknowledgements}
I thank S. Elitzur, E. Rabinovici and M. Smolkin for discussions and
useful comments.
\appendix
\section{Absence of tunneling}\label{Tunneling}
In this appendix we use our results in order to show that there is
no tunneling in the system if the global $O(N)$ symmetry is
unbroken.

As shown in section \ref{Quantum_case}, the dependence on time of
the expectation value of $\phi^2$ is governed by the classical EOM
with the effective potential (\ref{Quantum_Phi_eq_fixed_const}).
Therefore, if the field is at a minimum of the effective potential
the value of $<\phi^2>$ cannot change, and the tunneling is
impossible. In order to see this in a different way consider we
change our point of view and interpret our problem as a problem of a
motion of a particle of mass 1 in an $N$-dimensional space with a
spherically-symmetric potential $NU(r^2/N)$. Components of the field
$\vec{\phi}$ are interpreted as coordinates of the particle. We
consider a situation when the global $O(N)$ symmetry is unbroken,
which means that the wave function of the particle is spherically
symmetric (s-wave). The radial Schrodinger eq. is \be -\frac
12\,\frac{1}{r^{N-1}}\,\frac{d}{dr}\,r^{N-1}\,\frac{d}{dr}\,\Psi(r)
+NU\left(\frac{r^2}{N}\right)\,\Psi(r)=E\,\Psi(r)\,.\label{Radial_equation}\ee
Now make the following redefinitions: Define a radial wave function
as \be f(r)=r^{\frac{N-1}{2}}\,\Psi(r)\,,\ee introduce a new
variable $\xi=\frac{r}{\sqrt{N}}$ and assume that $N\gg 1$. The eq.
(\ref{Radial_equation}) becomes \be -\frac
{1}{2N^2}\frac{d^2}{d\xi^2}f(r)+\left[\frac{1}{8\,\xi^2}+
U\left(\xi^2\right)\right]\,f(\xi)=\frac{E}{N}\,f(\xi)\,.\ee This
eq. describes a motion of a particle of a large mass $N^2$ in the
effective potential (\ref{Quantum_effective_potential}). All
eigenvalues go to 0 as $N\to\infty$, hence the solutions of this eq.
have a small spread and therefore a tunneling is impossible.

Yet another way to see this is to carry out the instanton
computation.  The amplitude of tunneling is \be
\Gamma=A\,e^{-B}\,,\ee where $B$ is the action of a instanton and
the coefficient $A$ is proportional to a volume of the symmetry
manifold \cite{Banks:1973ps,Coleman:1977py}. In our case this
manifold is an $N-1$-dimensional sphere. Its volume is
$V=2\pi^{\frac{N-1}{2}}/\Gamma(\frac{N-1}{2})$, and as $N\to\infty$
the volume $V\to 0$. So the tunneling amplitude vanishes in this
limit.

\end{document}